# $10^{-7}$ contrast ratio at 4.5λ/D: New results obtained in laboratory experiments using nano-fabricated coronagraph and multi-Gaussian shaped pupil masks

**Abhijit Chakraborty and Laird A. Thompson**

*University of Illinois Urbana-Champaign, Astronomy Building, 1002 West Green Street, Urbana, IL 61801*
*abhijit@astro.uiuc.edu  thompson@astro.uiuc.edu*

**Michael Rogosky**

*Pennsylvania State University, 192 Materials Research Institute, University Park, PA 16802*
*mrogosky@engr.psu.edu*

**Abstract:**  We present here new experimental results on high contrast imaging of $10^{-7}$ at 4.5λ/D (λ=0.820 microns) by combining a circular focal plane mask (coronagraph) of 2.5λ/D diameter and a multi-Gaussian pupil plane mask. Both the masks were fabricated on very high surface quality (λ/30) BK7 optical substrates using nano-fabrication techniques of photolithography and metal lift-off. This process ensured that the shaped masks have a useable edge roughness better than λ/4 (rms error better than 0.2 microns), a specification that is necessary to realize the predicted theoretical limits of any mask design. Though a theoretical model predicts a contrast level of $10^{-12}$, the background noise of the observed images was speckle dominated which reduced the contrast level to $4 \times 10^{-7}$ at 4.5λ/D. The optical setup was built on the University of Illinois Seeing Improvement System (UnISIS) optics table which is at the Coude focus of the 2.5-m telescope of the Mt. Wilson Observatory.  We used a 0.820 micron laser source coupled with a 5 micron single-mode fiber to simulate an artificial star on the optical test bench of UnISIS.

©2005 Optical Society of America

---

## References and Links

1. Debes, John H.; Ge, Jian; Chakraborty, Abhijit, "First High-Contrast Imaging Using a Gaussian Aperture Pupil Mask," Astrophys. J. Lett. **572L**, 165-168, (2002)
2. Debes, John H.; Ge, Jian; Kuchner, Marc J.; Rogosky, Michael "Using Notch-Filter Masks for High-Contrast Imaging of Extrasolar Planets," Astrophys. J. **608**, 1095-1099, (2004)
3. Wilhelmsen Evans, Julia; Sommargren, Gary; Poyneer, Lisa; Macintosh, Bruce A.; Severson, Scott; Dillon, Daren; Sheinis, Andrew I.; Palmer, Dave; Kasdin, N. Jeremy; Olivier, Scot, "Extreme adaptive optics testbed: results and future work," Proc. SPIE **5490**, 954-959, (2004)
4. Kasdin, N. Jeremy; Vanderbei, Robert J.; Spergel, David N.; Littman, Michael G., "Extrasolar Planet Finding via Optimal Apodized-Pupil and Shaped-Pupil Coronagraphs," Astrophys. J. **582**, 1147-1161, (2003)
5. Vanderbei, Robert J.; Kasdin, N. Jeremy; Spergel, David N.,  "Checkerboard-Mask Coronagraphs for High-Contrast Imaging," Astrophys. J. **615**, 555-561, (2004)
6. Kuchner, Marc J.; Spergel, David N., "Notch-Filter Masks: Practical Image Masks for Planet-finding Coronagraphs, " Astrophys. J. **594**, 617-626, (2003)
7. Kuchner, M.J. , "A Unified View of Coronagraph Image Masks," preprint no. **astro-ph/0401256** at http://xxx.lanl.gov/archive/astro, (2004)
8. Debes, John H.; Ge, Jian, "High-Contrast Imaging with Gaussian Aperture Pupil Masks," PASP, **116**, 674-681, (2004)

9. Nakajima, T.; Oppenheimer, B. R.; Kulkarni, S. R.; Golimowski, D. A.; Matthews, K.; Durrance, S. T., "Discovery of a Cool Brown Dwarf," Nature **378**, 463-464, (1995)

10. Oppenheimer, Ben R.; Sivaramakrishnan, A.; Makidon, R.B., "Imaging Exoplanets: The role of small Telescopes," in *The Future of Small Telescopes in the New Millennium*, T. Oswalt, ed., **III**, (Dordrecht: Kluwar), p.157-174, (2003)

11. Oppenheimer, Ben R.; Digby, Andrew P.; Newburgh, Laura; Brenner, Douglas; Shara, Michael; Mey, Jacob; Mandeville, Charles; Makidon, Russell B.; Sivaramakrishnan, Anand; Soummer, Remi; and 7 coauthors, "The Lyot project: toward exoplanet imaging and spectroscopy," Proc. SPIE **5490**, 433-442, (2004)

12. Spergel, D.N., "A New Pupil for Detecting Extrasolar Planets," preprint no. **astro-ph/0101142** at http://xxx.lanl.gov/archive/astro, (2001)

13. Kuchner, Marc J.; Traub, Wesley A., "A Coronagraph with a Band-limited Mask for Finding Terrestrial Planets," Astrophys. J. **570**, 900-908, (2002)

14. Sivaramakrishnan, Anand; Lloyd, James P.; Hodge, Philip E.; Macintosh, Bruce A., "Speckle Decorrelation and Dynamic Range in Speckle Noise-limited Imaging," Astrophys. J. Lett. **581**, 59-62, (2002)

15. Thompson, Laird A.; Xiong, Yao-Heng, "Laser beacon system for the UnISIS adaptive optics system at the Mount Wilson 2.5-m telescope," Proc. SPIE **2534**, 38-47, (1995)

16. Thompson, Laird A.; Castle, Richard M.; Teare, Scott W.; McCullough, Peter R.; Crawford, Samuel L., "UnISIS: a laser-guided adaptive optics system for the Mt. Wilson 2.5-m telescope," Proc. SPIE **3353**, 282-289, (1998)

17. Thompson, Laird A.; Teare, Scott W., "Rayleigh Laser Guide Star Systems: Application to the University of Illinois Seeing Improvement System," PASP **114**, 1029-1042, (2002)

18. Thompson, Laird A.; Teare, Scott W.; Crawford, Samuel L.; Leach, Robert W., "Rayleigh Laser Guide Star Systems: UnISIS Bow-Tie Shutter and CCD39 Wavefront Camera," PASP **114**, 1143-1149, (2002)

19. Thompson, Laird A.; Teare, Scott W.; Xiong, Yao-Heng; Chakraborty, Abhijit; Gruendl, Robert, "Progress with UnISIS: a Rayleigh laser-guided adaptive optics system," Proc. SPIE **5490**, 90-96, (2004)

20. *Handbook of Microlithography, Micromachining and Microfabrication* Vol. 1, Rai-Choudhury, P., editor SPIE Optical Engineering Press ; London, UK : Institution of Electrical Engineers, (1997)

21. Sivaramakrishnan, Anand; Koresko, Christopher D.; Makidon, Russell B.; Berkefeld, Thomas; Kuchner, Marc J., "Ground-based Coronagraphy with High-order Adaptive Optics," Astrophys. J. **552** 397-408, (2001)

22. Trauger, John; Burrows, Chris; Gordon, Brian; Green, Joseph; Lowman, Andrew; Moody, Dwight; Niessner, Albert; Shi, Fang; Wilson, Daniel, "Coronagraph contrast demonstrations with the High Contrast Imaging Testbed," Proc. SPIE **5487**, 1330-1336, (2004)

# 1. Introduction

Direct optical/near-IR (λ = 0.5 to 2.2 microns) high contrast imaging at a dynamic range of $10^{-9}$ to $10^{-10}$ of nearby stars has recently emerged as one of the viable methods to detect extra-solar planets, thanks to recent work on how to control the diffraction pattern of star-light using focal plane masks (including classical coronagraphs), and spatially shaped or apodized pupil plane masks [1-12]. However, realizing contrast levels of $10^{-9}$ to $10^{-10}$ is proving to be a challenging task even in a laboratory environment with a laser source as the artificial star (see for eg. [3,22]). The major challenges are: a) fabricating masks with edge rms errors better than λ/4 in order minimize scattered light from the edges; b) minimizing scattered light from optical components in the detection system; and c) minimizing the rms error in the wavefront itself [13-14] to reduce the effects of speckles.

Here we report new results from lab experiments on high contrast imaging using a combination of a circular focal plane mask and a multi-Gaussian shaped pupil plane mask. These experiments were conducted on the optics table of the University of Illinois Seeing Improvement System (UnISIS). UnISIS is an adaptive optics (AO) system at the Coude focus of the 2.5-m Telescope of the Mt. Wilson Observatory (see [15-19]). We used the nano-fabrication technology of photolithography [20] to make both a circular spot for the focal plane mask and multi-Gaussian shaped pupil plane masks on extremely high quality BK7 optical substrates (λ/30 for λ=0.8 microns; 5 to 10 dig scratch or better). The shaped pupil mask was manufactured at the Pennsylvania State University nano-Fabrication facility while the circular spot focal plane mask was made by a commercial company. We describe the motivation and the coronagraph design in Section 2, the experimental setup and data recording in Section 3 and results and discussion in Section 4.

## 2. Motivation and design considerations for the shaped pupil mask and coronagraph for UnISIS

UnISIS is a high-order AO system with a recently upgraded control system that operates at an up-date rate as high as 1.6 KHz. Because we anticipate it to produce Strehl ratios of 0.5 – 0.6 under 1 arcesec seeing conditions (and even higher under sub-arcsec seeing conditions), we are motivated to assemble a high contrast imaging system to meet the following objectives: *1)* to achieve a dynamic range near bright stars that is ≤10⁻⁵ in the angular zone 4.5 to 6λ/D on the sky, *2)* to accommodate the wavelength range 0.8 microns to 1.65 microns, *3)* to work within the outer working angles of the UnISIS AO (12λ/D at 0.8 microns and ~6λ/D at 1.65 microns, see [10-11]), and *4)* to match the coronagraph to the 2.5-m telescope's large central secondary and its unusual tertiary mirror and secondary mirror support vanes.

For our first order system we considered a simple multi-Gaussian pupil mask that we inserted at a conjugate pupil plane (see Figs 3(a) and 3(b)). The first promising shaped pupil mask was designed by Spergel [12] and subsequently many others have appeared [1,4,8, and references therein]. For pupil masks with a simple Gaussian design, the top and bottom edges of the open areas are defined by Gaussian functions:

$$y_t = aR\{exp[-(\alpha x/R)^2] - exp(-\alpha^2)\}, \text{ and } y_b = -bR\{exp[-(\alpha x/R)^2] - exp(-\alpha^2)\} \tag{1}$$

where x goes from –R < x < R. The variables a, b, and α, are free parameters that can be optimized for a given telescope configuration. Since the Fourier transform of a Gaussian is another Gaussian, the intensity of the diffraction pattern along one axis in the image plane decreases exponentially. This is denoted as the high contrast axis. Thus variables a, b, and α can be used to optimize the aperture for depth of contrast, the inner working angle, and the azimuthal area of high contrast. For UnISIS the pupil mask was designed in such a way that it blocks the large central obscuration and the tertiary mirror support structure of the 2.5m telescope and at the same time maximizes the throughput (~20%) along two wide sectors in the focal plane for high contrast imaging. The only down side of this particular design is its somewhat high inner working angle of 4.5λ/D.

During the course of our work (see Sections 3 and 4 below), we recognized that both the number and the intensity of speckles was adversely affecting the dynamic range of the high contrast imaging. This could be caused by one of two effects: *(1)* the absence of dynamic wavefront error minimization in the final imaging optics (which all come after the UnISIS AO corrections) unlike [3, 22], or *(2)* that the bright star light was simply scattering from various optical elements including the edges of the pupil mask. We decided to address *point two* by introducing a hard-stop focal plane mask of 2.5λ/D that will affectively prevent the transmission of unneeded light from the central star fully recognizing that the scattered light from the hard stop might introduce its own subsidiary effects. Theory and design of hard-stop coronagraph systems with an optimized circular Lyot stop have been described by [21]. For a system with a hard focal plane stop, they define the ideal Lyot stop diameter as

$$L_d = D - (F) \cdot 2D/s \tag{2}$$

where *D* is the telescope diameter and s is the width of the focal plane mask in the natural units of λ/D. *F* is a free variable called the Lyot fine tuning parameter. Simulations reported in [21] show that simultaneously optimizing dynamic range and throughput leads to a number of *F* = 0.4 to 0.5 for an optical system similar to the Palomar or Gemini telescopes. For the 2.5-m telescope with its large central obscuration, we find an optimal solution with a focal plane mask of 2.5λ/D to require *F* values between 0.35 and 0.4. So in our unique situation, instead of a circular Lyot stop of constant *F* value, we have a multi-Gaussian shaped pupil mask whose *F* value varies between 0.25 and 0.45 but has an average value close to 0.35. It is likely that a focal plane mask of 3.5λ/D or 4λ/D in combination with the present Gaussian pupil mask may give even better performance, and we plan to fabricate and test such a mask in the future. In section 4 below, we present our simulations along with experimental results. Note that the present test-bed experiment with an artificial star source is a prelude to actual sky observations with UnISIS plus the final imaging optics that we test here.

## 3. Experimental setup for coronagraph and Gaussian shaped pupil plane mask on the UnISIS optics table

Figure 1(a) shows the design of the multi-Gaussian shaped pupil mask (GSPM) created on a BK7 optical substrate using photolithography. Figure 1(b) is a magnified photograph that shows the encircled part of the GSPM in (1a). The photograph clearly demonstrates the scale of the smoothness at the Gaussian edges which was found to be better than λ/4 (0.2 microns). However, the photograph also reveals pinhole defects introduced during photo-resist lift-off from the substrate. In order to get rid of these defects, another sample was made that did not have such defects within the central useable area with a 6.7mm diameter. This pupil mask is designed very specifically for the 6.7mm diameter conjugate telescope pupil plane on the UnISIS optics table.

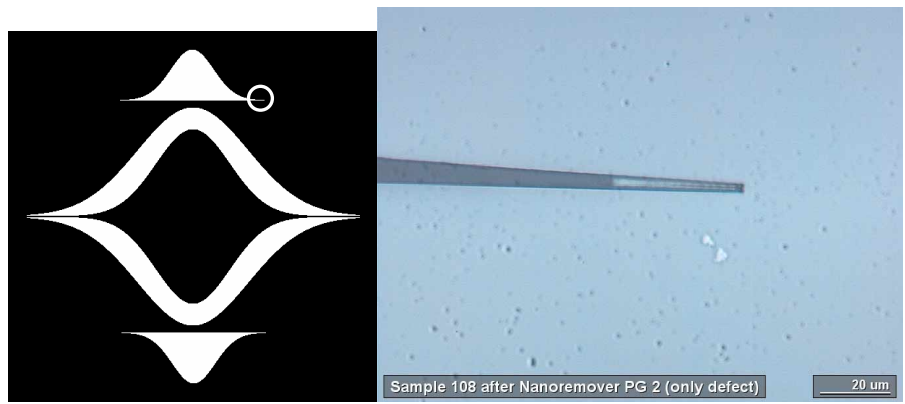

Fig. 1. (a). The design of the Gaussian mask. 1b). Enlarged high resolution image (taken at the nano-fab facility) of the encircled defective region in Fig. 1a showing the smoothness of the edges (note the scale of 20 microns; rms error on the edges was measured to be about 0.2microns) as well as pin-hole defects and residual metal at the Gaussian edge. Such defects were not present in the second sample which was used for the experiment.

Figure 2(a) shows the final pupil mask mounted in its holder. On the UnISIS optics table the test setup is arranged such that, with the help of a kinematically mounted folding mirror (M1, see Figs. 2(b) and 3), we have a choice of feeding light to the pupil mask with an artificial star (a single mode fiber source at 0.82 microns) for system characterization tests or feeding light to the pupil mask with a real on-sky star (UnISIS AO corrected star from the 2.5-m telescope). Our optical system also incorporates a 200 micron circular spot (2.5λ/D) as a focal plane mask (position I1 in Figs. 2(b) and 3) manufactured using the same process as that of the pupil mask but purchased commercially. Figures 2(b) and 3 show the optical setup on the UnISIS optics bench. It is worth mentioning here that the entire UnISIS optical bench is both thermally isolated and as well as protected from external scattered light with a tightly sealed cover. We also monitor the temperature difference across the optics table to minimize air flow and turbulence in the optical path during our experiments.

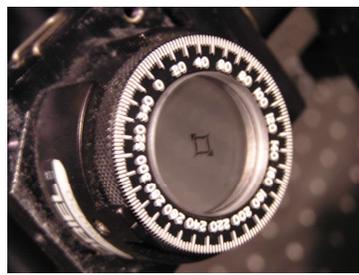
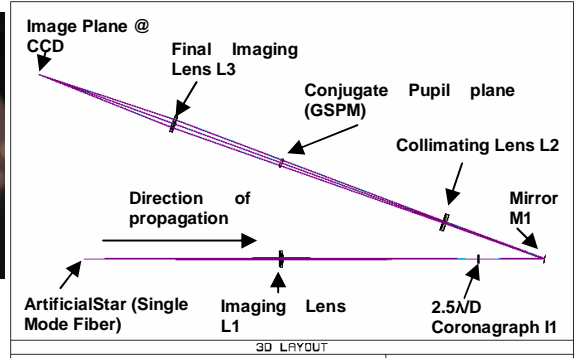

Fig. 2. (a) (left): Photograph showing the GSPM replicated on a 25 mm diameter BK7 substrate mounted in its holder. (b) (right): Sketch of the basic optical train of the test-bed set up shown in Fig. 3. Some of the elements that were not used (flipped) for recording the coronagraphic images are eliminated in this sketch, like the Pupil Re-Imaging Lens, the Dichroic Beam Splitter, Flip Mirror (M2), and the video camera (located on kinematic mounts between the Imaging Lens L3 and the CCD).

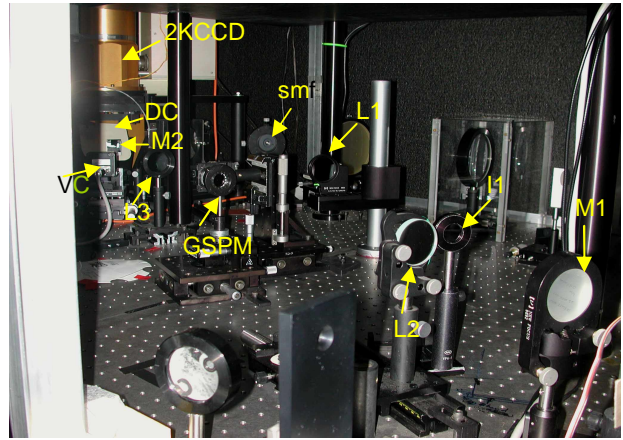

Fig. 3. Optical setup of the test-bench on the UnISIS AO optics table. **Legend:** Smf =Single mode fiber artificial star L1=Imaging Lens for image plane I1, I1=1st Image plane (for test-bench Coronagraph; focal plane mask), M1=Flip Mirror (for bending the beam), L2=Collimator Lens, GSPM=Gaussian Shaped Pupil Mask, L3=Final Imaging Lens onto the detector, DC=Dichroic Beam Splitter (50-50 between optical/NIR), M2=Flip Mirror 2, for redirecting the beam onto a video camera (VC) solely used for optical alignment purposes.

Diode laser light at 0.82 micron was fed into the 5-micron core of a single mode fiber using a standard coupler in such a way that the far (polished) end of the fiber served as an artificial star source for the lab experiment. All optical elements including the lenses used in the experiment have surface qualities of at least $\lambda/30$ at the 0.82 micron wavelength so the entire optical setup is diffraction limited (cf. Fig. 4(a)). Focal plane images were recorded with liquid nitrogen cooled 2048x2048 back illuminated Loral CCD with a quantum efficiency of 72% at the laser wavelength. The CCD has pixels 15 microns on a side and an rms read noise of 8.8 electrons. The effective plate scale on the detector with the present optics is about 6 pixels per resolution element ($\lambda/D$ @ $\lambda$=0.82 microns) or 0.015 arcsecs per pixel on the sky. Therefore the images are adequately sampled for high contrast imaging. Figure 4(a) shows a log stretched diffraction limited image of the artificial star before introducing the GSPM in the pupil plane and the focal plane mask in focal plane.

Figure 4(b) shows an image of the illuminated pupil. This pupil image was recorded by the CCD after introducing another lens identical to L1 between L3 and the CCD detector. (This additional lens is not shown in Fig. 3). The pupil re-imaging lens is mounted in such a way that it can be easily inserted and removed for pupil alignment. The sole purpose here is to align and focus the GSPM. We believe that the intensity variations (the features in grey) seen within GSPM are due to scattered light entering the camera at larger angles (> the f/81 imaging optical train) during this test with pupil re-imaging lens. This re-imaging lens is removed for the actual high contrast imaging experiment. Thus the diffraction limited image of a point source and the well illuminated and sharp pupil image, shown in Figs. 4(a) and 4(b) respectively, confirm the alignment of the optics. Finally, the focal plane mask was introduced at I1 for high contrast imaging.

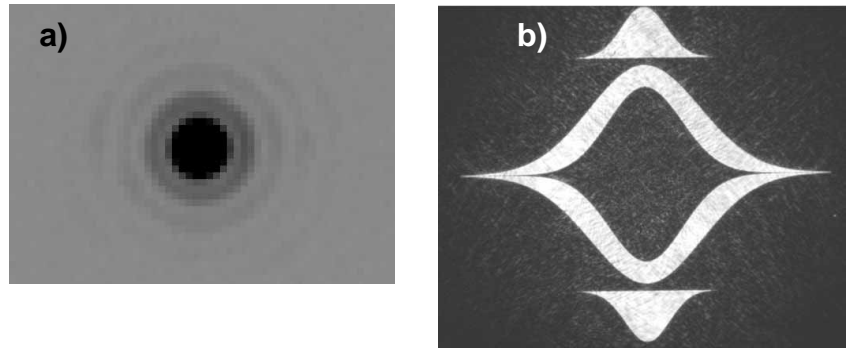

Fig. 4 (a) (left) Laser 0.82 micron point source log stretched negative image showing the Airy pattern before introducing the focal plane mask (coronagraph) at the focal plane and GSPM at the pupil plane. (b) (right) a positive log stretched image of the uniformly illuminated GSPM (see text for details). The pupil re-imaging lens has a wider field of view than the f/81 principal beam of the optical train. The intensity variations seen within the GSPM are probably due to light scattering and multiple reflections from the two surfaces of the BK7 substrate as well as light scattered at larger angles as seen by the pupil re-imaging lens.

The dynamic range of the CCD and its 16 bit ADC controller is $6.5 \times 10^4$. In order to obtain higher dynamic range images of $10^{-10}$ or greater, we obtained images at exposures times of 0.01s, 0.1s, 1s, 5s, 10s, 60s and 120s, and for each exposure time 9 images were recorded to provide better statistics. Thus a 0.01s exposure shows only the brightest parts of the coronagraphic image. 10s, 60s, or 120s images show the faintest signals along with the saturated regions. Given that the detector is linear, signals from the brighter regions can be estimated from the images of the shortest exposure times. Every image was bias subtracted, background subtracted and flat-fielded. All image analysis was done using the IRAF (Image Reductions Analysis Facility; see http://iraf.noao.edu) software package. Figure 5b shows the median of nine 10s images where a dynamic range of $10^{-9}$ was achieved, and the corresponding line profile across the "red" line is shown in Fig. 6. The pixels within $\pm 2\lambda/D$ from the centre are saturated due to spill-over of light from the edges of the coronagraph.

## 4. Results and discussion

Figure 5(a) shows the simulated model PSF for the case that contains both the $2.5\lambda/D$ circular focal plane mask and our GSPM in the pupil plane. The simulated model is constructed based on principles of Fraunhofer-diffraction (far-field approximation) in which the pupil and the focal plane are Fourier transforms of each other [21]. For the model calculation, we placed a uniformly illuminated circular telescope pupil (with no atmosphere above it) in square array filled with zeros. The square array was 6 times larger than the telescope aperture thus providing 6 samples over one resolution element ($\lambda/D$). We then followed the simulation procedures as described by [21]. The commercially available IDL software package was used for the PSF model calculation. The analogous experimental configuration is shown in Fig. 5(b).

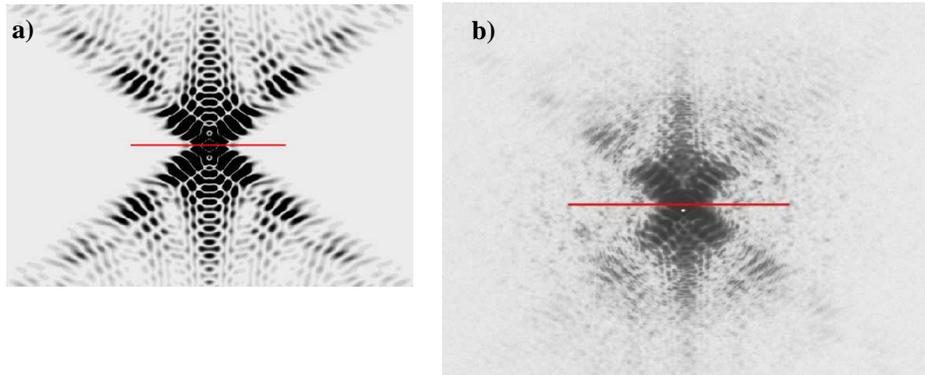

Fig. 5. (a) (left) Theoretically simulated model PSF with 2.5λ/D focal plane mask and the GSPM (negative image). (b) (right) The observed PSF (log stretched negative image) on the UnISIS optics table using 0.82 micron laser source coupled with a 5 micron single mode fiber as an artificial star. Note that the high contrast region is along the "red" line in the images. The observed PSF is the median of nine 10s images (in (b)). The image shows many speckles (tiny back dots) in the high contrast region making the background speckle-noise limited rather than detector-noise limited. See text for details.

A qualitative comparison between the two PSF images shows striking similarities although the background noise in the observed PSF was found to be significantly higher and speckle dominated. Line profiles across the "red" lines in the images are shown in Fig. 6. The contrast levels quoted here are normalized with respect to the un-obscured artificial star (occulting focal plane mask removed) and with our GSPM at the Lyot plane. If we ignore the region inside a radius of 4.5λ/D, the high points in the experimental profile scatter around a contrast level of $4x10^{-7}$ while the valleys scatter around a contrast level of $8x10^{-8}$. The total dynamic range in the image was $10^{-9}$ so our experimental setup met the challenge. Since the noise in the image was dominated by speckles at contrast levels of $10^{-7}$ to $10^{-8}$, there was no need to increase the dynamic range beyond $10^{-9}$ in this lab experiment. In the near future with the help of a larger focal plane mask of 3.5λ/D or 4λ/D it may be possible to lower the number and intensity of the speckles. Note that from the images recorded using GSPM alone we found a dynamic range of $6x10^{-6}$ at 5.5 - 6λ/D. Both the number and the intensity of speckles were significantly lower once the hard focal plane mask was added (see section 2).

The other published results that approach similar contrast levels in coronagraphic or pupil plane mask imaging are those of [3] and some what better by [22]. The aims of these authors were somewhat different from ours. In Ref. [3] the experimental bench-mounted optics had no focal plane obscuration and used a prolate spheroidal pupil mask manufactured by laser cutting, and yet in ref. [22] there was a precisely made apodized coronagraph. Both had optical systems that incorporated a deformable mirror in the imaging section which made it possible to interactively reduce wavefront errors to something below 1.8 nm rms. By doing so, [3] reached a dynamic range of $8x10^{-8}$ (peaks) to $2x10^{-8}$ (valleys) beyond 8λ/D and [22] reached $3x10^{-9}$ at 4λ/D. We feel gratified to obtain a dynamic range similar to that reported in [3] but at radial distances as close as 4.5λ/D to our test star. In fact, without any simple means to further correct the wavefront errors in the as-built UnISIS coronagraph which we have tested here, any "non-common path" aberrations (i.e., those not controllable with the UnISIS deformable mirror) could hamper our efforts to successfully complete high contrast imaging. Thus the results from the present experiment provide an upper limit in terms of a dynamic range that can be achieved with the UnISIS final imaging optics, a limit that surpasses our initial goal of $\leq 10^{-6}$ at $5\lambda/D$ with real sky observations. These results provide an excellent prognosis for future success on the sky with the UnISIS coronagraph.

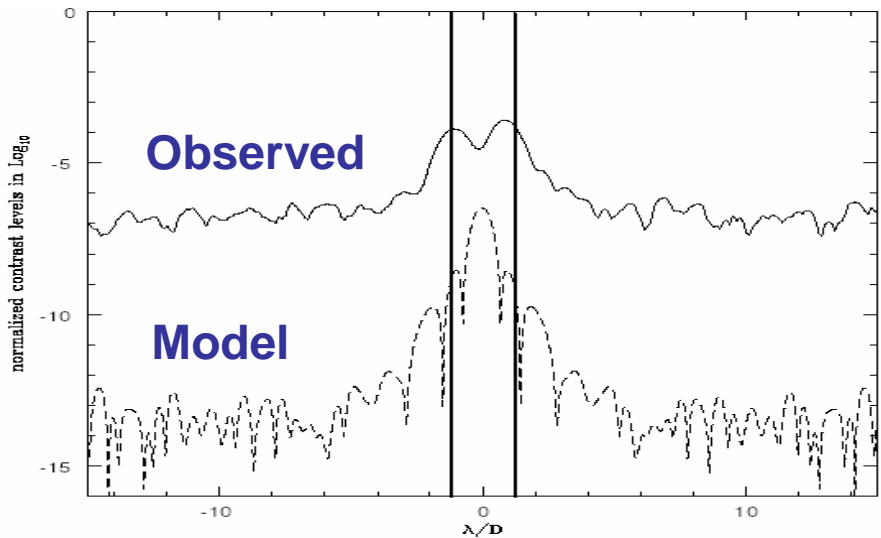

Fig. 6. A comparison of line profiles showing normalized contrast levels of observed and model PSF across the "red" line in the high contrast region (see Figs. 5(a) and 5(b). The difference between the two is mainly due to speckle noise in the observed PSF line profile. The y-axis represents the normalized dynamic range or the contrast level (normalized with respect to the peak intensity of the un-obscured test star with the GSPM at the Lyot plane). The two bold solid lines at $\pm 1.25\lambda/D$ show the width of the $2.5\lambda/D$ focal plane mask. Also, between $\pm 2\lambda/D$ the observed line profile is saturated due to spill-over of scattered light over the edges of the focal plane mask.

One additional step yet to be included in [3, 22] and in our work is PSF subtraction. In other words, if two high contrast images taken with the same coronagraph are subtracted from each other, many residual non-common path errors can be eliminated. If we work on the sky in this manner, we will have a coronagraph that is, by itself, capable of pushing toward a contrast level in the range of $5\times10^{-8}$. Long before we reach this level with observations of objects on the sky, undoubtedly we will encounter other troublesome barriers associated with the details of obtaining high Strehl AO corrected images. Extreme AO is a discipline that, no doubt, will require persistence in beating back a whole series of barriers in terms of achievable contrast ratios. The test of our coronagraph described here is just the first step.

A comparison with an optimized classical coronagraph [21] shows that, in this case, in the final image plane residues of the circular diffraction pattern remain at contrast levels of $10^{-5}$ to $10^{-6}$ with respect to their un-occulted test star at $\sim6\lambda/D$ (and poorer contrast levels at smaller radii). We simulated a classical optimized coronagraph with a $2.5\lambda/D$ focal plane stop, 99.9% Strehl and $F$=0.5, and our simulation reached $\sim5\times10^{-6}$ at $5\lambda/D$. Such circular residual patterns are absent along the high contrast axis in the case of the GSPM. From our model calculations of the combined $2.5\lambda/D$ circular focal plane mask and GSPM, the background residues are at $10^{-12}$ contrast levels, thus significantly enhancing the theoretical limit of the dynamic range. In the real world, the "*hybrid coronagraph system*" (focal plane mask + GSPM) will have the added advantage of blocking scattered star-light in the optical train, but its benefits will be realized only at higher Strehls ($\geq70\%$).

In a broader context, we would like to emphasize the advantages of the coronagraph configuration we have tested here, namely a $2.5\lambda/D$ on-axis hard spot in the focal plane and a GSPM in the pupil plane. If the focal and pupil plane masks are manufactured with the exquisitely precise techniques of nanofabrication, then it is likely that the best ground-based Extreme AO systems as well as precision space-based telescopes will be able to explore the interesting regions in the near vicinity of stars like imaging disks around young stars and searching for substellar dwarfs as faint companions. The regions around bright galaxy nuclei will also be interesting as point sources and jets in these regions (and possible disk-like features) will be detectable with the best optical configurations.

**Acknowledgments**

This material is based upon work supported by the National Science Foundation under Grant No. AST-0096741. Any opinions, findings, and conclusions or recommendations expressed in this material are those of the authors and do not necessarily reflect the views of NSF. The authors would like to thank the Mt. Wilson Observatory staff for their assistance during these tests and the Penn State University nano-fabrication facility staff for the GSPM fabrication. AC would like to thank John Debes for useful discussions during the fabrication of the pupil mask. The authors would like to thank the anonymous reviewers for their useful comments which helped to improve the quality of the paper.